\documentclass[preprint]{sig-alternate-05-2015}
\usepackage[noadjust]{cite}

\usepackage{times} 
\usepackage[USenglish]{babel}
\usepackage{xcolor}
\usepackage{listings}
\usepackage{braket}
\definecolor{darkblue}{rgb}{0.0, 0.0, 0.55}
\usepackage[colorlinks=true,allcolors=black,breaklinks,draft=false]{hyperref}   
\usepackage[final]{microtype}
\usepackage{paralist} 
\usepackage{booktabs}
\usepackage{tabularx}
\usepackage{stmaryrd}
\usepackage{amsmath}
\usepackage{amssymb}  
\usepackage{graphicx}
\usepackage{makecell}

\usepackage{amsthm}
\usepackage{thmtools}

\usepackage[keeplastbox]{flushend}

\usepackage{algorithm}
\usepackage[noend]{algpseudocode}    
\usepackage{bussproofs}
\usepackage{varwidth}
\usepackage{tikz}
\usepackage{stackengine}
\usetikzlibrary{arrows, shadows, calc, matrix, decorations.pathmorphing}
\usepackage{pifont}

\usepackage{pgfplots}
\usepackage{pgfplotstable}
\pgfplotsset{compat=1.5}

\usepackage{xpatch}
\xapptocmd\normalsize{%
 \abovedisplayskip=5pt plus 3pt minus 9pt
 \abovedisplayshortskip=0pt plus 3pt
 \belowdisplayskip=5pt plus 3pt minus 9pt
 \belowdisplayshortskip=1pt plus 3pt minus 4pt
}{}{}

\newenvironment{mathprooftree}
  {\varwidth{.9\textwidth}\centering\leavevmode}
  {\DisplayProof\endvarwidth}

\lstdefinelanguage{Lambda}{%
  morekeywords={%
    if,let,in,then,else,where,fix 
  },%
  morekeywords={[2]int},   
  otherkeywords={:}, 
  literate={
    {->}{{$\to$}}{2}
    {lambda}{{$\lambda$}}{1}
  },
  basicstyle={\scriptsize\sffamily},
  keywordstyle={\scriptsize\bfseries},
  keywordstyle={[2]\scriptsize\itshape}, 
  keepspaces,
}[keywords,comments,strings]%

\newcommand{\mycomment}[1]{}

\makeatletter
\let\OldStatex\Statex
\renewcommand{\Statex}[1][3]{%
  \setlength\@tempdima{\algorithmicindent}%
  \OldStatex\hskip\dimexpr#1\@tempdima\relax}
\makeatother

\newcommand{\textcode}[1]{\texttt{#1}}

\makeatletter
\def\@copyrightspace{\relax}
\makeatother

\algrenewcommand\algorithmicindent{0.7em}%

\theoremstyle{definition}

\begin{document}
\toappear{}

\title{Flow-Sensitive Composition of Thread-Modular\\ Abstract Interpretation}

\numberofauthors{2}
\author{
\alignauthor
Markus Kusano \\
  \affaddr{Virginia Tech}\\
  \affaddr{Blacksburg, VA, USA}\\
\alignauthor
Chao Wang \\
  \affaddr{University~of~Southern~California}\\
  \affaddr{Los Angeles, CA,  USA}\\
}

\maketitle

\begin{abstract}

We propose a constraint-based flow-sensitive static analysis for
concurrent programs by iteratively composing thread-modular abstract
interpreters via the use of a system of lightweight constraints.  Our
method is compositional in that it first applies sequential abstract
interpreters to individual threads and then composes their results.
It is flow-sensitive in that the causality ordering of interferences
(flow of data from global writes to reads) is modeled by a system of
constraints.  These interference constraints are lightweight since
they only refer to the execution order of program statements as
opposed to their numerical properties: they can be decided efficiently
using an off-the-shelf Datalog engine.  Our new method has the
advantage of being more accurate than existing, flow-insensitive,
static analyzers while remaining scalable and providing the expected
soundness and termination guarantees even for programs with unbounded
data.  We implemented our method and evaluated it on a large number of
benchmarks, demonstrating its effectiveness at increasing the accuracy
of thread-modular abstract interpretation.

\end{abstract}

%
%

%
%

%
%



\section{Introduction}

Although abstract interpretation~\cite{Cousot77} has wide use in the
analysis and verification of sequential programs, designing a scalable
abstract-interpretation-based analysis for shared-memory concurrent
programs remains a difficult task~\cite{Ferrara08, Mine11, Mine12,
  Mine14, Farzan12}.  Due to the large concurrent state space,
directly applying techniques designed for sequential abstract
interpretation to interleaved executions of a concurrent program does
not scale.  In contrast, recent thread-modular
techniques~\cite{Ferrara08, Mine11, Mine12, Mine14} drastically
\emph{over-approximate} the interactions between threads, allowing a
more tractable but less accurate analysis.  Their main advantage is
that sequential abstract interpreters can be lifted to concurrent ones
with minimal effort.  However, they consider thread interactions in a
\emph{flow-insensitive} manner: given a system of threads $\{A,B,C\}$,
for instance, they assume $A$ can observe all combinations of
memory modifications from $B$ and $C$ despite that some of these
combinations are infeasible, thereby leading to a large number of
false alarms even for simple programs.

In this paper, we propose the first constraint-based flow-sensitive
method for composing sequential abstract interpreters to form a more
accurate thread-modular analysis.  
%
%
Though desirable, no existing static method is able to maintain inter-thread
flow sensitivity with a reasonable cost.
The main advantage of our method is that, through the use of a lightweight
system of constraints, it can achieve a high degree of flow sensitivity with
negligible runtime cost.  
Here,
our goal is to prove the correctness of reachability
properties of a program: the properties are embedded assertion
statements whose error conditions are relational expressions over
program variables at specific thread locations.  Another advantage is
that our method can be implemented as a flexible composition of
existing sequential abstract-interpretation frameworks while retaining
the well-known benefits such as soundness and guaranteed termination as well as the
freedom to plug in a large number of abstract domains~\cite{Cousot77,
  Mine06}.

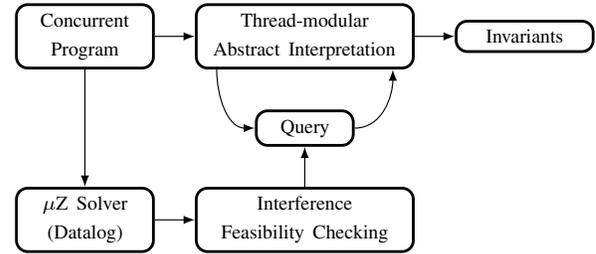
\begin{figure}
  \centering
  \resizebox{.95\linewidth}{!}{%
  \begin{tikzpicture}
    \matrix (m) [
      matrix of nodes
      , column sep=5mm
      , row sep=5mm
       , ampersand replacement=\&
      , nodes={draw 
        , line width=1pt
        , anchor=center 
        , text centered
        , rounded corners
      }
      , txt/.style={minimum width=1cm,anchor=center},
      ] 
      {
           |[txt, text width=1.5cm]| {\scriptsize Concurrent\\ Program}                       
         \& |[txt, text width=2.5cm]| {\scriptsize Thread-modular\\ Abstract Interpretation}    
         \& |[txt, text width=1.5cm]| {\scriptsize Invariants}                             
         \\
         \& |[txt, text width=1cm]| {\scriptsize  Query} 
         \&
         \\
          |[txt, text width=1.5cm]|  {\scriptsize  $\mu${Z} Solver (Datalog)}      
         \& |[txt, text width=2.5cm]| {\scriptsize  Interference\\ Feasibility Checking} 
         \\
      }; 
      \draw[-latex] (m-1-1.east) -- (m-1-2.west);
      \draw[-latex] (m-1-2.east) -- (m-1-3.west);

      \draw[-latex] ([xshift=0.00cm]m-2-2.east) to [out=0,in=270]  ([xshift=1.10cm]m-1-2.south);
      \draw[-latex] ([xshift=-1.10cm]m-1-2.south) to [out=270,in=180]  ([xshift=-0.00cm]m-2-2.west);

      \draw[-latex] (m-1-1.south) -- (m-3-1.north);
      \draw[-latex] (m-3-1.east) -- (m-3-2.west);
      \draw[-latex] (m-3-2.north) -- (m-2-2.south);

  \end{tikzpicture}
  } 
  \caption{\textsc{Watts}: Flow-sensitive thread-modular analysis.}
  \label{fig:block}
\end{figure}

Figure~\ref{fig:block} shows an overview of our new method.  Given 
a concurrent program, our method returns a set of relational and
numerical invariants statically computed at each thread location as
output.  These invariants, in turn, can be used to prove the set of
reachability properties of the program.  During the thread-modular
analysis, we first apply a sequential abstract interpreter to each
individual thread and then propagate their results across threads
before applying these sequential abstract interpreters again.  The
iterative process continues until a fix-point is reached over the set
of invariants.  During each iteration, the abstract interpreter also
communicates with a Datalog engine to check if a thread
interference, or set of interferences (data flow from global writes to
reads), is feasible.  If we can statically prove that the interference
is infeasible, i.e., it cannot occur in any real execution of the
program, we skip it, thereby reducing the analysis time and increasing
accuracy.

In contrast to existing methods in this domain, our
analysis is \emph{flow-sensitive} for two reasons.
First, we explore the memory interactions between threads individually
by propagating their memory-states along data-flow edges without
eagerly merging them through join operations as in prior
techniques~\cite{Ferrara08, Mine11, Mine12, Mine14}.  Second, we
identify and remove the infeasible memory interactions by constructing
and solving a system of \emph{lightweight} happens-before 
constraints.  These constraints (Horn clauses in finite domains)
capture only the causality ordering of the program's statements as
opposed to complex relational/numerical properties.  As such, they can
be solved by a Datalog engine in polynomial time.  These two
techniques, together, greatly reduce the number of false alarms
caused by over-approximating the global memory state across threads,
thereby allowing more properties to be verified compared to prior
approaches.

Consider the program in Figure~\ref{fig:ex01}, which has two threads
communicating through the shared variables \textcode{x} and
\textcode{flag}. Initially \textcode{flag} is false and \textcode{x}
is 0. Thread~1 only performs shared memory writes by setting
\textcode{x} to 4, and then to 5, before setting \textcode{flag} to
true. Thread~2 only performs shared memory reads: it reads the value
of \textcode{flag} and if the value is true, reads the value of
\textcode{x}. Note that the \textcode{ERROR!} (at $l_{13}$) is
unreachable since, for Thread~2 to reach $l_{11}$, Thread~1 has to set
\textcode{flag} to true (at $l_6$) before $l_9$ is executed; but in
such a case, $l_5$ must have been executed, meaning \textcode{x} must
have been set to \textcode{5}.

\begin{figure}
\vspace{1ex}
  \centering
  \begin{minipage}{0.35\linewidth}
    \begin{lstlisting}[language=C, numbers=left, frame=single, name=motiv1]
bool flag = false;
int x = 0;
void thread1() {
  x = 4;
  x = 5;
  flag = true;
}
    \end{lstlisting}
  \end{minipage}
  \hspace{0.15\linewidth}
  \begin{minipage}{0.35\linewidth}
    \begin{lstlisting}[language=C, numbers=left, frame=single, name=motiv1]
void thread2() {
  bool b1 = flag;
  if (b1) {
    int t1 = x;
    if (t1 != 5) 
      ERROR!;
} }
    \end{lstlisting}
  \end{minipage}
  \caption{Proving the \textcode{ERROR!} on $l_{13}$ is not reachable.}
  \label{fig:ex01}
\end{figure}

Prior static analyzers such as Ferrara~\cite{Ferrara08} and
Min{\'e}~\cite{Mine11,Mine12,Mine14} would have difficulty because
their treatment of inter-thread communication is
\emph{flow-insensitive}.  That is, if one thread writes to a shared
variable at program location $l_i$ and another thread reads the same
shared variable at $l_j$, they model the interaction by adding a
data-flow edge from $l_i$ to $l_j$ even if the edge is infeasible or
is only feasible in some program executions.  For example, in
Figure~\ref{fig:ex01}, no concrete execution simultaneously allows the
flow of \textcode{x} from $l_4$ to $l_{11}$ and the flow of
\textcode{flag} from $l_6$ to $l_{9}$.  In such cases, these prior
methods would lose accuracy because their way of modeling the
inter-thread data flow cannot differentiate between the feasible and
infeasible data-flow combinations.

In contrast, our new method detects and eliminates such infeasible
data-flows.  For now, it suffices to say that our method would report
that the flow of \textcode{x} from $l_4$ to $l_{11}$ \emph{cannot
  co-exist} with the flow of \textcode{flag} from $l_6$ to $l_9$.  We
will provide full details of our constraint-based interference
analysis in Section~\ref{sec:constr}.

Our method for checking the feasibility of
inter-thread data flows is sound: when it declares a certain
combination of interferences as infeasible, the combination is
guaranteed to be infeasible.  However, for efficiency reasons, our
method does not attempt to identify every infeasible combination.
This is consistent with the fact that abstract interpretation, in the
context of property verification, is generally an over-approximation:
it can prove the \emph{absence} of errors but does not aim to
guarantee that all unverified properties have real violations.  As
such, the additional effort we put into our constraint-based
interference analysis is a fair trade-off between lower runtime
overhead and improved accuracy.  This puts our method in a nice middle
ground between the more heavyweight model checkers~\cite{Clarke04b}
and the more scalable and yet less accurate static 
analysis techniques~\cite{Ferrara08, Mine11, Mine12, Mine14}.

Another perhaps subtle benefit of our method is that the sequential
abstract interpreter only needs a lightweight constraint
solver~\cite{Hoder11} as a black-box to query the feasibility of a set
of interferences.  As such, it provides a flexible and extensible
framework, allowing additional constraints, deduction rules, and
decision procedures (e.g., solvers for symbolic-numerical domains) to
be plugged in to further reduce the number of false alarms.  To make
our method more efficient, we also propose several optimizations to
our interference feasibility analysis (Section~\ref{sec:pruning}): we
leverage control and data dependencies to group interferences before
checking the feasibility of their combinations, and leverage
property-directed pruning to reduce the program's state space.

Our method differs from the \textsc{Duet} concurrent static analyzer
of Farzan and Kincaid~\cite{Farzan12,FarzanK13} despite that both
methods employ constraint-based analysis, since we aim at solving a
different problem.  First, our goal is to accurately analyze a
concurrent program with a fixed number of threads, whereas their goal
is to soundly approximate the behavior of a parameterized program with
an unbounded number of thread instances. Second, our method is
strictly thread-modular: we iteratively apply a sequential abstract
interpreter to a set of control-flow graphs, one per thread, and one
at a time.  In contrast, they analyze a single monolithic data-flow
graph of the entire concurrent program. As a result, their method is
significantly less accurate than ours on non-parameterized programs.
We illustrate the main difference between these two computational
models, i.e., a set of per-thread control-flow graphs versus a
monolithic data-flow graph, in Section~\ref{sec:motiv:duet}.

We implemented our method in a static analysis tool named
\textsc{Watts}, for verifying reachability properties of multithreaded
C/C++ programs written using the POSIX thread library.  The tool
builds upon the LLVM compiler, using the $\mu${Z}~\cite{Hoder11}
fix-point engine in Z3~\cite{DeMoura08} to solve Datalog constraints
and the Apron library~\cite{Jeannet09} to implement the sequential
abstract interpreter over numerical abstract domains.  We have
evaluated our method on a set of benchmarks with a total of 26,309
lines of code.  Our experiments show that \textsc{Watts} can
successfully prove 1,078 reachability properties, compared to 526
properties proved by \textsc{Duet}~\cite{Farzan12,FarzanK13} and 38
properties proved by the prior, flow-insensitive
methods~\cite{Mine12,Mine14}.  Furthermore, \textsc{Watts} achieved
a 28x increase in the number of verified properties with only a 1.4x
increase in the analysis time.

In summary, this paper makes the following contributions:
\begin{compactenum}
  \item We propose a flow-sensitive method for composing
    thread-modular abstract interpreters into a more accurate static
    analysis procedure.
  \item We develop a lightweight constraint-based framework for
    soundly checking the feasibility of inter-thread interferences and
    combinations of interferences.
  \item We develop optimization techniques to improve the efficiency
    of our analysis by leveraging control and data dependencies and
    property-directed pruning.
  \item We implement and evaluate our method on a large set of
    benchmarks and demonstrate its advantages over prior works.
\end{compactenum}


\section{Motivating Examples}
\label{sec:motiv}

We present a series of examples showing applications of our new method compared
to existing approaches.

\subsection{Thread-modular Abstract Interpretation}

First, Figure~\ref{fig:ex01} provides an
overview of prior works on thread-modular abstract
interpretation~\cite{Ferrara08, Mine11, Mine12, Mine14}.  These
methods all use the same notion of \emph{interference} between
threads: an interference is a value stored into shared memory at some
point during the execution of a thread. In Figure~\ref{fig:ex01},
there are three interferences, all from Thread~1: the writes to
\textcode{x} at $l_4$ and $l_5$ and the write to \textcode{flag} at
$l_6$.

These prior techniques analyze the program by statically computing the
over-approximated set of interferences for each thread:
\begin{compactenum}
  \item Initially, the set of interferences in each thread is empty.
  \item Each thread is independently analyzed in the presence of interferences
    from all other threads.
  \item The set of interferences in each thread is recomputed based on the 
    results of the analysis in Step 2.
  \item Steps 2--3 are repeated until the interferences stabilize.
\end{compactenum}

\noindent
During the thread-modular analysis (step 2), each thread keeps track
of its own \emph{memory environment} at every thread location. The
memory environment is an abstract state mapping program variables to
their values.  To incorporate inter-thread effects, when a thread
performs a shared memory read on some global variable \textcode{q}, it
reads either the values of \textcode{q} in its own memory environment,
or the values of \textcode{q} from the interferences of all other
threads.  These techniques rely on a flow-insensitive analysis in that
each read may see \emph{all} values ever written by any other thread,
even if the flow of data is not feasible in all, or any, of the
concrete program executions.

\begin{table}
  \caption{Running prior approaches~\cite{Ferrara08, Mine11, Mine12, Mine14} on Figure~\ref{fig:ex01}.
}
  \label{tbl:ex01Prior}

  \centering
  \resizebox*{\linewidth}{!}{%
  \begin{tabular}{*{5}{c}}
    \toprule
    & \multicolumn{2}{c}{Thread 1} & \multicolumn{2}{c}{Thread 2} \\
    \cmidrule(r){2-3} \cmidrule(r){4-5}
    Iteration & ~Reachable~ & Interference        & ~Reachable~        & Interference \\
    \midrule
    One       
      & 4,5,6  
      & \makecell{$\mathit{flag} = \{1\}$ \\ $x=\{4, 5\}$}
      & 9,10         
      & $\varnothing$ \\
    \midrule
    Two       
      & 4,5,6  
      & \makecell{$\mathit{flag} = \{1\}$ \\ $x=\{4, 5\}$}
      & \makecell{9,10,11 \\ 12,13}
      & $\varnothing$ \\
    \bottomrule
  \end{tabular}
}
\end{table}

Table~\ref{tbl:ex01Prior} shows the results of analyzing
Figure~\ref{fig:ex01} with prior thread-modular
approaches~\cite{Ferrara08, Mine11, Mine12, Mine14}.  Column 1 shows
the two iterations.  Columns 2 and 4 show the lines reachable after
each iteration in the two threads. Columns 3 and 5 show the interferences generated after
each iteration.  In the second iteration, the interferences generated
during the first iteration are visible: Thread~2 is analyzed in the
presence of the interferences generated by Thread~1.  After two
iterations, the interferences stabilize, which concludes the analysis.
Unfortunately, the result in Table~\ref{tbl:ex01Prior} shows that
Thread~2 can reach $l_{12}$, where it reads the value of \textcode{x}
either from its own memory environment (the initial value
\textcode{0}) or from the interference of Thread~1 (\textcode{4}),
thereby allowing the \textcode{ERROR!} to be reached.  This is a false
alarm: the property violation is generated because the inter-thread
interferences are handled in a flow-insensitive manner.

%
%
In this example, to eliminate the false alarm one has to maintain a complex
invariant such as $(\mathit{flag}=\mathit{true}) \rightarrow (x = 5)$ which
cannot be expressed precisely as a relational invariant even in expensive
numerical domains such as convex polyhedra.
Additionally, in order to propagate such a relational invariant across threads,
as in~\cite{Mine14}, they need to hold over all states within a thread.
Otherwise, interference propagation is inherently non-relational.
Specifically, propagating the interferences on a variable \textcode{x} first
requires a projection on \textcode{x}, thus forgetting all relational
invariants.

In contrast, our method can eliminate the false alarm even while staying in
inexpensive abstract domains such as intervals.
%
%
In particular, our work shows that eagerly joining over all interference across
threads is inaccurate and should be avoided as much as possible.

\subsection{Iterative Flow-sensitive Analysis}

We propose, instead, to partition the set of interferences from other
threads into clusters and then consider combinations of interferences
only within these clusters. In this way, we effectively delay the join
of interferences and avoid the inaccuracies caused by eagerly
joining in existing methods.  For example, if we assume the three
interferences in Figure~\ref{fig:ex01} fall into one cluster (worst
for efficiency but best for accuracy), our analysis of the program
would be as follows: in the first iteration, we apply per thread
abstract interpretation and then compute the interferences for each
thread; these computations remain the same as in the first iteration
of Table~\ref{tbl:ex01Prior}.  In the second iteration, however, when
analyzing Thread~2 at the point of reading \textcode{flag}, there will
be six possible cases, due to the Cartesian product of $x=\{0,4,5\}$
and $\mathit{flag}=\{0,1\}$.

Unlike prior approaches, which eagerly join these cases to form $x =
\{0,4,5\} \land \mathit{flag} = \{1,0\}$, we analyze the impact of
each case $\rho_1$--$\rho_6$ \emph{individually} as follows:
\begin{compactitem}
  \item $\rho_1$, corresponding to $(x = 4 \land \mathit{flag} = 0)$;
  \item $\rho_2$, corresponding to $(x = 5 \land \mathit{flag} = 0)$;
  \item $\rho_3$, corresponding to $(x = 0 \land \mathit{flag} = 0)$; 
  \item $\rho_4$, corresponding to $(x = 5 \land \mathit{flag} = 1)$;
  \item $\rho_5$, corresponding to $(x = 4 \land \mathit{flag} = 1)$; and
  \item $\rho_6$, corresponding to $(x = 0 \land \mathit{flag} = 1)$.
\end{compactitem}
\noindent
This leads to enough accuracy to prove the \textcode{ERROR!}
is not reachable.  First, when $\mathit{flag} = 0$ ($\rho_1$,
$\rho_2$, and $\rho_3$) the \textcode{ERROR!} cannot be reached since the
branch at $l_{10}$ will not be taken (\textcode{b1} is
\textcode{false}).  Second, in the case of $\rho_4$, the first branch
at $l_{10}$ will be taken but the branch guarding \textcode{ERROR!}
will not, since $x = 5$, meaning \textcode{t1} is also 5.  For the two
remaining cases ($\rho_5$ and $\rho_6$) our constraint-based
interference analysis (Section~\ref{sec:constr}) would show it is
impossible to have both $x=4 \land \mathit{flag}=1$, or $x=0 \land \mathit{flag} = 1$.

The intuition behind the analysis is that infeasible data flows cause
a contradiction between program-order constraints and data-flow edges.
Specifically, examining $\rho_5$, if Line~9 reads \textcode{flag} as
\textcode{1} and Line~11 reads \textcode{x} as \textcode{4}, then:
\begin{compactitem}
\item Line 6 is executed before Line 9 (\textcode{b1==true}),
\item Line 9 is executed before Line 11 (program order),
\item Line 11 is executed before Line 5 (\textcode{t1==4}), and
\item Line 5 is executed before Line 6 (program order).
\end{compactitem}
\noindent
This leads to a contradiction since the above must-happen-before
relationship forms a cycle, meaning the combination cannot happen.
Similarly, $\rho_6$ is infeasible since the write of 1 to
\textcode{flag} implies the updates to \textcode{x} have already
occurred, meaning \textcode{x}'s initial value, 0, is not visible to
Thread 2. At this point, the only feasible interferences do not cause
an \textcode{ERROR!}  --- the program is verified.

To obtain the aforementioned accuracy, we leverage the statically
computed control and data dependencies to partition the set of
interferences into clusters.  This can significantly reduce the number
of cases considered during our thread-modular analysis.  For example,
when a load of \textcode{y} is \emph{independent} of the subsequent
load of \textcode{x}, e.g., the value loaded from \textcode{y} has no
effect on the load of \textcode{x}, the thread would have two
unconnected subgraphs in its program dependence
graph~\cite{Ferrante87}.  Unconnected subgraphs create a natural
partition of loads into clusters, thereby significantly reducing the
complexity of our interference-feasibility checking.  This is because
we only need to consider combinations of interferences \emph{within}
each subgraph.  We will show details of this optimization in
Section~\ref{sec:pruning}.

\subsection{Control-flow versus Data-flow Graphs}
\label{sec:motiv:duet}

Our method also differs from \textsc{Duet}, a concurrent static
analyzer for parametric programs~\cite{Farzan12,FarzanK13}.  Although
\textsc{Duet} also employs a constraint-based analysis, its
verification problem is significantly different.  First, it is
designed for soundly analyzing parameterized concurrent programs,
where each thread routine may have an unbounded number of instances.
In contrast, our method is designed to analyze programs with a fixed
number of threads with the goal of obtaining more accurate analysis
results.

Second, \textsc{Duet} relies on running an abstract interpreter over a
single data-flow graph of the entire program, whereas our method
relies on running abstract interpreters over a set of thread-local
control-flow graphs.  The difference between using a set of
thread-local control-flow graphs and a single monolithic data-flow
graph can be illustrated by the following two-threaded program:
\textcode{\{x++;\} || \{tmp=x;\}}.
In the monolithic data-flow graph
representation~\cite{Farzan12,FarzanK13}, there would be cyclic
data-flow edges between the read and write of \textcode{x} across
threads as well as an edge from the write of \textcode{x} to itself.
As a result, applying a standard abstract interpretation based
analysis would lead to the inclusion of $\mathit{tmp}=\infty$ as a possible
value, despite that in any concrete execution of the program, the end
result is either $\mathit{tmp}=1$ or $\mathit{tmp}=0$ (assume that $x=0$
initially). Our method, in contrast, can correctly handle this
program.

\section{Background}
\label{sec:background}

We provide a brief review of abstract interpretation based static
analysis for sequential and concurrent programs.  For a thorough
treatment, refer to Nielson and Nielson~\cite{Nielson99} and
Min{\'e}~\cite{Mine11, Mine12, Mine14}.

\subsection{Sequential Abstract Interpretation}

An abstract interpretation based static analysis is a fix-point
computation in some abstract domain over a program's
\emph{control-flow graph} (CFG).  The CFG consists of nodes
representing program statements and edges indicating transfer of
control between nodes.  Due to their one-to-one mapping we
interchangeably use the term statement and node.  We assume the graph
has a unique entry.

The analysis is parameterized by an \emph{abstract domain} defining
the representation of \emph{environments} in the program.  An
environment is an abstract memory state.  The purpose of restricting
the representation of memory states to an abstract domain is to reduce
computational overhead and guarantee termination.  For example, in the
interval domain~\cite{Cousot77}, each variable has an upper and lower
bound.  For a program with two variables \textcode{x} and
\textcode{y}, an example environment is $x = [0, 5] \land y = [10,
  20]$.  With properly defined meet ($\sqcap$) and join ($\sqcup$)
operators, a partial order ($\sqsupseteq$), as well as the top
($\top$) and bottom ($\bot$) elements, the set of all possible
environments in the program forms a lattice.  In the interval domain,
for example, we have $[0, 5] \sqcup [10, 20] = [0, 20]$ and $[0, 5]
\sqsupseteq [0, 2]$.

Each statement in the program is associated with a \emph{transfer
  function}, taking an environment as input and returning a new
environment as output. The transfer function of statement $st$ for
some input environment $e$ returns a new environment $e'$, which is
the result of applying $st$ in $e$.  Consider the above example of
interval domain for $x$ and $y$ again.  The result of executing the
statement \textcode{x=x+y} in the above example environment would
be the new environment $x = [10, 25] \land y = [10, 20]$.

For brevity, we will not define all the transfer functions for a
programming language explicitly since the main contributions of this
work are language-agnostic.  As an example, however, consider the
statement \textcode{t=load x}, which copies a value from memory to a
variable.  Its transfer function can be represented as $\lambda e. e
[ t = x ]$, where $e [ st ]$ is the
result of evaluating $st$ in the environment $e$.  Conceptually, it
takes an input environment and returns a new environment where $t$ is
assigned the current value of $x$.

\begin{algorithm}
  \caption{Sequential abstract interpretation.}
  \label{alg:seq_ai}
  {\footnotesize
    \begin{algorithmic}[1]
      \Function{SeqAbsInt}{ $G$ : the control-flow graph }
      \State $\mathit{Env(n)}$ is initialized to $\top$ if $n\in\Call{Entry}{G}$, else to $\bot$
        \State $\mathit{WL} \gets \Call{Entry}{G}$
        \While { $\exists n \in \mathit{WL}$ }
        \State $\mathit{WL} \gets \mathit{WL}\setminus \{n\}$
          \State $e \gets \Call{Transfer}{n, \mathit{Env}(n)}$
          \ForAll {$n' \in \Call{Succs}{G, n}$ such that $e \not \sqsubseteq \mathit{Env}(n')$}
              \State $\mathit{Env}(n') \gets \mathit{Env}(n') \sqcup e$
              \State $\mathit{WL} \gets \mathit{WL}\cup \{n'\}$
          \EndFor
        \EndWhile
        \State \Return $\mathit{Env}$
      \EndFunction
    \end{algorithmic}
  }
\end{algorithm}

The standard work-list implementation of an abstract-interpretation
based analysis~\cite{Nielson99} is shown in
Algorithm~\ref{alg:seq_ai}.  The input is a control-flow graph $G$,
where \textsc{Entry}($G$) is the entry node and \textsc{Succs}($G,n$)
is the set of successors of node $n$.  $\mathit{Env}$ is a function
mapping each node $n$ to an environment immediately before $n$ is
executed.  The initial environment $\top$ associated with the entry
node means that all program variables can take arbitrary values, e.g.,
$x = y = \cdots = [-\infty,\infty]$ for integer variables.  The
initial environments for all other nodes are set to $\bot$ (the
absence of values).

The work-list, $\mathit{WL}$, is initially populated only with the
entry node of the control-flow graph. The fix-point computation in
Algorithm~\ref{alg:seq_ai} is performed in the while-loop: a node
$n\in \mathit{WL}$ is removed and has its transfer function executed,
resulting in the new environment $e$.  The function \textsc{Transfer}
takes a node $n$ and the environment $Env(n)$ as input and returns the
new environment $e$ (result of executing $n$ in $Env(n)$) as output.
If a successor of the node $n$ has a current environment with less
information than $e$ (as determined by $\not\sqsubseteq$), then it is
added to the work-list and its environment is expanded to include the
new information (Lines 7-9). The process proceeds until the work-list
is empty, i.e., all the environments have stabilized.  Standard
widening and narrowing operators~\cite{Cousot77} may be used at Line~8
to guarantee termination and ensure speedy convergence.

\subsection{Thread-modular Abstract Interpretation}
\label{sec:tm-ai}

Next, we review thread-modular abstract interpretation: an iterative
application of a sequential abstract interpreter on each thread in the
presence of a joined set of interferences from all other threads.
Since a thread-modular analysis never constructs the \emph{product}
graph of all threads in the program, it avoids the state space
explosion encountered by non-thread-modular methods~\cite{Jeannet13}.

First, we make a slight modification to the previously described
sequential abstract interpretation (Algorithm~\ref{alg:seq_ai}); the
\emph{per-thread} abstract interpretation must consider both the
thread-local environment and the \emph{interferences} from other
threads.  Here, an interference is an environment resulting from
executing a shared memory write.  Let
\textsc{SeqAbsInt-Modified}$(G,i)$ be the modified abstract analyzer,
which takes an additional environment $i$ as input.  The environment
$i$ represents a joined set of interferences from all the other
threads.  We also modify the transfer function
\textsc{Transfer}$(n,Env(n))$ of shared memory read as follows: for
\textcode{t=load x}, where \textcode{x} is a shared variable, we allow
\textcode{t} to read either from the thread-local environment $Env(n)$
or from $i$, the interference parameter.  For example, if the
thread-local environment before the load statement contains $x = [10,
  15]$ and the interference parameter contains $x = [50, 60]$, we
would have $t = [10, 15] \sqcup [50, 60] = [10, 60]$.

\begin{algorithm}
  \caption{Thread-modular abstract interpretation.}
  \label{alg:tm_ai}
  {\footnotesize
    \begin{algorithmic}[1]
      \Function{ThreadModAbsInt}{ $\mathit{Gs}$ : the set of CFGs }
        \State $\mathit{TE} \gets \varnothing$
        \State $I \gets \varnothing$
        \Repeat
          \State $I' \gets I$
          \ForAll {$g \in \mathit{Gs}$}
            \State $i \gets \bigsqcup \{e \mid e\in I(g'),  g'\in \mathit{Gs}, \text{ and } g'\neq g \}$
                   \Comment{Sec.~\ref{sec:tm-ai}}
            \State $\mathit{Env} \gets \Call{SeqAbsInt-Modified}{g, i}$
            \State $\mathit{TE} \gets \mathit{TE} \uplus \mathit{Env}$
          \EndFor

        \ForAll {$(n, e) \in \mathit{TE}$}
          \If {$n$ is a shared memory write in $g\in \mathit{Gs}$}
            \State $I(g) \gets I(g) \sqcup \Call{Transfer}{n, e}$
          \EndIf
        \EndFor

        \Until{ $I = I'$ }
        \State \Return $\mathit{TE}$
      \EndFunction
    \end{algorithmic}
  }
\end{algorithm}

Algorithm~\ref{alg:tm_ai} shows the thread-modular analysis procedure.
The input is the set $\mathit{Gs}$ of control-flow graphs, one per
thread.  The output, $\mathit{TE}$, is a function mapping the thread
nodes (nodes in all threads) to environments.  During the analysis,
each thread-local CFG $g$ has an associated interference environment
$I(g)$: the environment is the join of all environments produced by
shared memory writes in the thread $g$.  Due to their one-to-one
correspondence, we will use thread and its (control-flow) graph
interchangeably.

Inside the thread-modular analysis procedure, both $\mathit{TE}$ and
$I$ are initially empty.  Then, the sequential abstract interpretation
procedure is invoked to analyze each thread $g\in Gs$.  The
environment $i$ (Line~7) is the join of all interfering environments
from other threads.  The sequential analysis result, $\mathit{Env}$,
is a function mapping nodes in $g$ to their corresponding
environments.  With a slight change of notation, we use $\mathit{TE}
\uplus \mathit{Env}$ (Line~9) to denote the join of environments from
$TE$ and $Env$ on their matching nodes. Let $A$ and $B$ be sets of
pairs of the form $\{(n,e),\ldots\}$; then $A \uplus B$ denotes the
join of environments on the matching nodes.

After analyzing all the threads (Lines~6--9), we take the results
($\mathit{TE}$) and compute the new interferences: for each thread
$g$, the new environment $I(g)$ is the join of all environments
produced by the shared memory writes (Lines~10--12).  The analysis
repeats until the interferences stabilize ($I = I'$),
meaning that environments in all node ($\mathit{TE}$) also
stabilize.  Again, standard widening and narrowing
operators~\cite{Cousot77} may be used to ensure speedy convergence.
Overall, the thread-modular analysis is an additional fix-point
computation on the set of interferences relative to sequential
analysis, with the same termination and soundness
guarantees~\cite{Mine14}.

\section{Flow-sensitive Thread-modular Analysis}
\label{sec:iterative}

In this section, we present our new method for flow-sensitive
thread-modular analysis.  For ease of comprehension, we shall postpone
the presentation of the constraint-based feasibility checking until
Section~\ref{sec:constr}, while focusing on explaining our method for
maintaining inter-thread flow-sensitivity during thread-modular
analysis.

\subsection{The New Algorithm}

Before diving into the new algorithm, notice that the reason why
Algorithm~\ref{alg:tm_ai} is flow-insensitive is because all
environments from interfering stores of other threads are joined
(Line~7) prior to the thread-modular analysis.  Furthermore, within
the thread-modular analysis routine, \textsc{SeqAbsInt-Modified}, the
combined interfering environment, $i$, is joined again with the
thread-local environment during the application of the transfer
function at each CFG node.  Such eager join operations are the main
sources of inaccuracy in existing methods.  First, inaccuracy arises
from the join operation itself: it tends to introduce additional
behaviors, e.g., $[0, 0] \sqcup [10,10] = [0, 10]$.  Second, a thread
is allowed to see \emph{any} combination of interfering stores even if
some of them are obviously infeasible (e.g., Section~\ref{sec:motiv},
Figure~\ref{fig:ex01}).


To avoid such drastic losses in accuracy, we need to make fundamental
changes to the thread-modular analysis procedure.
\begin{asparaitem}
\item 
For each thread $g \in \mathit{Gs}$, instead of defining  its
interference as a single environment, we use a set of
pairs $(n,e)$ where $n$ is a CFG node of a shared memory write and $e$
is the environment after $n$.
\item
For each shared variable read, instead of it reading from the eagerly
joined set of environments, we maintain a set, $\mathit{LIs}(l) =
\{(n,e),\ldots \}$, where each $(n,e)$ represents an interfering store
and the store's interfering environment.
\item 
For each thread $g\in Gs$, instead of representing the interferences
from all other threads as the join of the interfering environments
(Line~7, Algorithm~\ref{alg:tm_ai}), we represent them as a set $I_c$
of \emph{interference combinations}: each $i_c\in I_c$ is a distinct
combination of the store-to-load flows for all $l\in\Call{Loads}{g}$.
\end{asparaitem}

Algorithm~\ref{alg:constr_ai} shows our new analysis: in the remainder
of this section, we shall compare it with Algorithm~\ref{alg:tm_ai}
and highlight their differences.  There are two main differences.
First, the interferences are represented as a set of pairs of store
statements and their associated environment (Line~13).  We modify
$\uplus$ to be the join of environments of pairs with matching nodes
across two sets. Recall that if $A$ and $B$ are sets of pairs of the
form $\{(n,e),\ldots\}$, then $A \uplus B$ denotes the join of
environments on the matching nodes.  Second, we compute the set $I_c$
of feasible and non-redundant interference combinations (store-to-load
flows) for a thread (Line~7) and analyze a thread in the presence of
each combination individually (Lines~8--10).  That is, for each call
to the sequential abstract interpreter \textsc{SeqAbsInt-Modified2},
as the second parameter, instead of passing the join of interferences
from all other threads, we pass each $i_c\in I_c$ to map every load to
an interfering store individually.

\begin{algorithm}[!t]
  \caption{Flow-sensitive thread-modular analysis.}
  \label{alg:constr_ai}
  {\footnotesize
    \begin{algorithmic}[1]
      \Function{ThreadModAbsInt-Flow}{$\mathit{Gs}$: the set of CFGs}
        \State $\mathit{TE} \gets \varnothing$
        \State $I \gets \varnothing$
        \Repeat
          \State $I' \gets I$
          \ForAll {$g \in \mathit{Gs}$}
            \State \textcolor{black}{$\mathit{I_c} \gets \Call{InterferenceComboFeasible}{g,\mathit{I}}$}
            \ForAll {\textcolor{black}{ $i_c \in \mathit{I_c}$ }}
                    \Comment{Sec.~\ref{sec:iterative}}
                    \State $\mathit{Env} \gets \Call{SeqAbsInt-Modified2}{g, \textcolor{black}{i_c}}$
                 \State $\mathit{TE} \gets \mathit{TE} \uplus \mathit{Env}$
            \EndFor
          \EndFor

          \ForAll {$(n, e) \in \mathit{TE}$}
            \If {$n$ is a shared memory write in $g\in \mathit{Gs}$}
               \State $I(g) \gets I(g) \uplus \{ \Call{Transfer}{n,e} \}$
            \EndIf
          \EndFor

        \Until{$I = I'$}
        \State \Return $\mathit{TE}$
      \EndFunction

      \State {~~}

      \Function{InterferenceComboFeasible}{$g$, $I$}
        \State $I_c \gets \varnothing$
        \State $\mathit{VEs} \gets \{(n,e)~|~ (n,e)\in I(g'),~ g'\in \mathit{Gs}, \text{ and } g'\neq g \}$

        \ForAll {$l\in\Call{Loads}{g}$}
          \State $\mathit{LIs}(l) \gets \{ (s_{\mathit{dummy}}, e_{\mathit{self}}) \}$
          \If {$l$ is not self-reachable}
            \ForAll { $(n,e) \in \mathit{VEs}$ }
              \If { $\Call{LoadVar}{l} = \Call{StoreVar}{n}$}
                \State $\mathit{LIs}(l) \gets \mathit{LIs}(l) \cup \{ (n,e) \}$
              \EndIf
            \EndFor
          \Else \Comment{Handling loads in loops}
            \ForAll { $(n,e) \in \mathit{VEs}$ }
              \If { $(\Call{LoadVar}{l} = \Call{StoreVar}{n})$
                    \Statex[6] $\phantom{e} \land \lnot
                  \Call{MustHappenBefore}{l, n}$}
                \State $\mathit{LIs}(l) \gets \mathit{LIs}(l) \uplus \{ (s_{\mathit{dummy}},e) \}$
              \EndIf
            \EndFor
          \EndIf
        \EndFor

        \State $\mathit{Es} \gets \Call{CartesianProduct}{\mathit{LIs}}$   
                \Comment{Sec.~\ref{sec:pruning}}
        \ForAll {$i_c \in \mathit{Es}$}
           \If {\textcolor{black}{$\Call{Query.IsFeasible}{i_c}$}}      
                \Comment{Sec.~\ref{sec:constr}}
              \State $I_c \gets I_c \cup \{ i_c \}$
           \EndIf
        \EndFor
        \State \Return $\mathit{I_c}$
      \EndFunction

    \end{algorithmic}
  }
\end{algorithm}

\subsection{The Interference Combinations}

Inside \textsc{InterferenceComboFeasible}($g,I$), we compute the set
$I_c$ of feasible interference combinations.  Here, $\Call{Loads}{g}$
is the set of shared variable reads in thread $g$, $\Call{LoadVar}{l}$
is the variable used in the load instruction $l$, and
$\Call{StoreVar}{s}$ is the variable stored-to in the store
instruction $s$.

We first compute the set $\mathit{VEs}$ of interferences from other
threads (Line~19); each pair $(n,e)\in VEs$ is a store and environment
from a thread other than $g$.  Then, we pair each load
$l\in\Call{Loads}{g}$ with any corresponding store in $\mathit{VEs}$
(Lines~20--29); the result is stored in $\mathit{LIs}$ which maps each
load instruction $l$ to a set of stores in the form of $(n,e)$ pairs.
The special pair $(s_\mathit{dummy}, e_\mathit{self})$ indicates the
thread should read from its intra-thread environment.  For now, ignore
Lines~26--29 since they are related to the handling of loops --- we
discuss how loops are handled during the computation of interference
combinations in the next subsection.

Next, the function $\textsc{CartesianProduct}$ takes $\mathit{LIs}$ as
input and returns the complete set of interference combinations from
$\mathit{LIs}(l_1)\times\cdots \times \mathit{LIs}(l_k)$.  To make
what we have explained so far clearer, consider an example program
with two threads: $g_1$ and $g_2$.  Thread $g_1$ has two loads,
$\Call{Loads}{g_1}=\{l_1,l_2\}$ such that $\Call{LoadVar}{l_1} =
\textcode{x}$ and $\Call{LoadVar}{l_2} = \textcode{y}$.  Thread $g_2$
has three interfering environments: two on \textcode{x}, $s_1$ and
$s_2$, with associated environments $e_1$ and $e_2$, respectively; and
another, $s_3$, on \textcode{y}, with environment $e_3$.  Assume we
are currently analyzing $g_1$ in the presence of interferences from
$g_2$.

We first use the set $I$ of interferences to collect the interferences
from $g_2$ in $\mathit{VEs}$: $\{(s_1, e_1), (s_2, e_2), (s_3,
e_3)\}$.  Next, we compute $\mathit{LIs}$ for the two loads
$\{l_1,l_2\}$ in thread $g_1$.  We pair $l_1$ with the two
interferences on \textcode{x} from $s_1$ and $s_2$, and pair $l_2$
with the single interference on \textcode{y} from $s_3$.  Using
$[\cdots]$ to denote a list of items, we represent the result as
$\mathit{LIs}(l_1) = [ (s_1, e_1), (s_2, e_2), (s_\mathit{dummy},e_\mathit{self}) ]$ and
$\mathit{LIs}(l_2) = [ (s_3,  e_3), (s_\mathit{dummy},e_\mathit{self}) ]$.
Without any optimizations, the resulting Cartesian product
$\mathit{Es} = \mathit{LIs}(l_1) \times \mathit{LIs}(l_2)$ would
contain the following items:
\[\begin{array}{ll}
i_{c_1} = \{ \langle l_1,(s_1,e_1) \rangle,                              & \langle l_2,(s_3,e_3) \rangle \}, \\
i_{c_2} = \{ \langle l_1,(s_2,e_2) \rangle,                              & \langle l_2,(s_3,e_3) \rangle \}, \\
i_{c_3} = \{ \langle l_1,(s_{\mathit{dummy}},e_{\mathit{self}})\rangle,  & \langle l_2,(s_3,e_3)\rangle \},\\
i_{c_4} = \{ \langle l_1,(s_1,e_1) \rangle,                              & \langle l_2,(s_{\mathit{dummy}},e_{\mathit{self}}) \rangle \},\\
i_{c_5} = \{ \langle l_1,(s_2,e_2) \rangle,                              & \langle l_2,(s_{\mathit{dummy}},e_{\mathit{self}}) \rangle \},\\
i_{c_6} = \{ \langle l_1,(s_{\mathit{dummy}},e_{\mathit{self}}) \rangle, & \langle l_2,(s_{\mathit{dummy}},e_{\mathit{self}})\rangle \}.
\end{array}\]
For each combination $i_c\in \mathit{Es}$, we check if it is feasible
(Lines~31--33): the infeasible combinations will be filtered out, and
the result, $I_c$, is returned.  We discuss how we determine the
feasibility of an interference (Line~32) in Section~\ref{sec:constr}.

Continuing with the algorithm's description, on Line~9 the sequential
abstract interpretation, \textsc{SeqAbsInt-Modified2}, takes $g$ and
each $i_c\in I_c$ as input and returns a node-to-environment map,
$\mathit{Env}$, as output.  During this per-thread analysis, the
transfer function of a load uses only $i_c$ to determine the
environment to use.  When a load $l_1$ is being executed, if the
special item $\langle l_1,
(s_{\mathit{dummy}},e_{\mathit{self}})\rangle$ is in $i_c$, the load
reads from its own thread-local environment at $l_1$; if the remote
store environment $\langle l_1,(s, e)\rangle$ is in $i_c$, the load
also reads from the remote environment $e$.

At this point, we have improved the prior work
(Algorithm~\ref{alg:tm_ai}) to avoid inaccuracies from
over-approximations caused by the eager join over all interferences.
The cost for this accuracy is explicitly testing each of the
combinations of potential interferences.  However, we have not
presented our methods for clustering and pruning
(Section~\ref{sec:pruning}) as well as checking if any of the
combinations are \emph{infeasible} (Section~\ref{sec:constr}).  By
applying such optimization techniques, we cannot only drastically
reduce the overhead of running the abstract interpretation subroutine
but also increase the accuracy.

\subsection{Handling Loops}
\label{sec:loops}

Since a load within a loop may execute many times, the number of
stores it could read from may be infinite.  To guarantee termination,
we join all the interfering stores that \emph{may} affect a load in a
loop with the environment within the thread at the time of the load.
By doing this, we conservatively treat all these feasible
interferences in a flow-insensitive manner for loads within loops.

Specifically, Lines 26--29 perform the join of interferences for loads
within a loop. For a given load, all stores on the same variable that
must-not-happen after the load are considered (we will further discuss
the happens-before constraints in Section~\ref{sec:constr}). For these
conflicting stores, all of the environments are joined together on a
single dummy node ($s_{\mathit{dummy}}$). In the end, each
self-reachable load has a single (joined) environment. Consequently,
during the Cartesian product computation, it will have a single
interference. Within the sequential abstract interpreter, the load
merges the thread-local environment and this single interfering
environment.

However, even in such case, our new method is more accurate than the
prior work.  Consider the example in Figure~\ref{fig:loop}. Thread 1
executes a load in a while-loop running an arbitrary number of times
concurrently with thread 2 before creating thread 3. Because of the
thread creation, there is a must-happen-before edge between the load
in thread 1 (Line 5) and the write in thread 3 to \textcode{x}. When
constructing the interference combinations for the load in thread 1
($l$), there are three potential stores: $s_{10}$, $s_{11}$, and
$s_{14}$ for the writes to \textcode{x} on Lines 10, 11, and 14,
respectively.

When considering $s_{10}$, the condition on Line 28 of our new
algorithm is true since $s_{10}$ does not always happen after $l$ (and
similarly for $s_{11}$).  Therefore, $\mathit{LIs}(l)$ is assigned
$\{s_{\mathit{dummy}}, e_{10}\}$ initially, where $e_{10}$ is the
environment at $s_{10}$.  Next, $\mathit{LIs}(l)$ is assigned
$\{s_{\mathit{dummy}}, e_{10} \sqcup e_{11}\}$.  Finally, for
$s_{14}$, since it must happen after $l$, it is not added to
$\mathit{LIs}$.  When computing the Cartesian product, there is only a
single load with a single location-store pair, so there is only one
interference combination.

For this example, the analysis results in \textcode{t1} being 0, 1, or
2.  The value of $10$ written by thread 3 is excluded using the
must-happen-before constraint.  So, although multiple interfering
stores are merged for the single load within the loop, the accuracy of
the analysis is still higher than prior flow-insensitive analyses.

\begin{figure}
\vspace{1ex}
  \centering
  \begin{minipage}{0.35\linewidth}
    \begin{lstlisting}[language=C, numbers=left, frame=single, name=loop]
int x = 0;
void thread1() {
  create(thread2);
  while (*) {
    int t1 = x;
  }
  create(thread3);
}
    \end{lstlisting}
  \end{minipage}
  \hspace{0.15\linewidth}
  \begin{minipage}{0.35\linewidth}
    \begin{lstlisting}[language=C, numbers=left, frame=single, name=loop]
void thread2() {
  x = 1;
  x = 2;
}

void thread3() {
  x = 10;
}
    \end{lstlisting}
  \end{minipage}
  \caption{Example: handling loops in thread-modular analysis.}
  \label{fig:loop}
\end{figure}

\subsection{Correctness}
\label{sec:proofs}

Our method in Algorithm~\ref{alg:constr_ai} is a form of semantic
reduction~\cite{Sagiv99, Cousot79} of the interferences allowed by the
prior flow-insensitive approach in
Algorithm~\ref{alg:tm_ai}. Specifically, the input environment to a
load instruction in Algorithm~\ref{alg:tm_ai} is the join of the set
$S = \{\rho, \rho_1, \ldots, \rho_n\}$ where $\rho$ is the
intra-thread environment and $\rho_1, \ldots, \rho_n$ are environments
from interfering stores.  The semantic-reduction operator we use in
Algorithm~\ref{alg:constr_ai} is to apply the transfer function of the
load to each element of $S$ individually relative to all other loads
(i.e., the Cartesian product). Therefore, the correctness of our
algorithm directly follows the correctness argument in
\cite{Sagiv99,Cousot79}.  Additionally, we remove infeasible
interferences combinations (Lines 31-33), which does not affect the
soundness of the algorithm.


In the case of loops, the transfer function of a load can be executed
more than once: each execution of the transfer function may use a
different interference, so, using the same semantic-reduction operator
would have resulted in a potentially infinite number of interference
combinations.  In this case, we conservatively merge all the
\emph{feasible} interferences into a single value. Correctness of this
treatment directly follows the correctness of
Algorithm~\ref{alg:tm_ai}.

In the case of aliasing, our algorithm can be lifted to use the output
of any (sound) alias analysis by considering each alias-set as a
single variable -- it is a standard technique to handle aliasing in
static analysis.  In such case, our algorithm would operate on these
alias-sets instead of on the individual program variables.

\section{Constraint-based Feasibility}
\label{sec:constr}

We now present our procedure for eliminating infeasible
combinations of interferences.  We revisit
Algorithm~\ref{alg:constr_ai} to show its integration with our new
thread-modular analysis procedure.

Removing infeasible interferences from the thread-modular analysis
significantly reduces computational overhead and increases accuracy.
However, the main problem is that the feasibility checking has to be
conducted efficiently for such an optimization to be useful.
Therefore, our goal is to make the checking both \emph{sound} and
\emph{efficient}.  By sound, we mean that if the procedure determines
a combination is infeasible then it is truly infeasible.  By
efficient, we mean that the procedure relies on constructing and
solving a system of \emph{lightweight} constraints, i.e., Horn clauses
in finite domains, which can be decided using a Datalog engine in
polynomial time.

\begin{algorithm}
  \caption{Constraint-based feasibility checking.}
  \label{alg:constr_check}
  {\footnotesize
    \begin{algorithmic}[1]
      \State $\mathit{POs} \gets \Call{ProgramOrder-Constraints}{\mathit{Gs}}$
      \State $\Call{Query.Add}{\mathit{POs}}$
      \Function{Query.IsFeasible}{$i_c$: permutation of interferences}
        \State $\mathit{Cs} \gets \Call{ReadsFrom-Constraints}{i_c}$
        \State $\Call{Query.Add}{\mathit{Cs}}$
        \State $\mathit{res} \gets \Call{Query.Satisfiable}$
        \State $\Call{Query.Remove}{\mathit{Cs}}$
        \State \Return $res$
      \EndFunction
    \end{algorithmic}
  }
\end{algorithm}

Algorithm~\ref{alg:constr_check} shows the high-level flow of our
feasibility analysis procedure.  Initially, we traverse the set $\mathit{Gs}$
of control-flow graphs to compute a set $\mathit{POs}$ of constraints
representing the order between statements
which must hold on all possible executions of the program.
We initialize the constraint system with these orderings by calling
$\Call{Query.Add}{\mathit{POs}}$.  

During the execution of Algorithm~\ref{alg:constr_ai} (Lines~31--33),
for each $i_c\in I_c$, we compute a set $\mathit{Cs}$ of \emph{reads-from}
constraints, which must be enforced in order to realize the interference
combination $i_c$. We add them to the system as well by calling
$\Call{Query.Add}{\mathit{Cs}}$.

Our constraint analysis then, using a set of deduction rules, expands
upon these input constraints to generate more constraints.  We invoke
$\textsc{Query.Satisfiable}$ to check if the constraint system is
satisfiable.  The deduction rules are designed such that, if the
system is not satisfiable, then $i_c$ is
guaranteed to be infeasible.
In the remainder of this section, we go into each of these steps in
detail.

\subsection{The Program-order and the Reads-from Constraints}

To check the simultaneous feasibility of $\mathit{POs}$ and
$\mathit{Cs}$, we first compute the \emph{dominators} on a thread's
CFG.  Given two nodes $m$ and $n$ in a graph $g$, $m$
dominates $n$ if all paths from the entry of $g$ to $n$ go through
$m$.  Then, we define the following relations:
\begin{asparaitem}

  \item $\textsc{Dominates}$ is the dominance relation on a 
    thread's CFG: $(m,n)\in \textsc{Dominates}$ means  $m$
    dominates  $n$.
  
  \item $\textsc{NotReachableFrom}$ is reachability on a thread's CFG:
    $(m,n) \in \textsc{NotReachableFrom}$ means node $m$ can not be reached from node $n$.

  \item $\textsc{ThCreates}$ is a parent--child relation over threads: $(p,n_\mathit{sta}) \in \textsc{ThCreates}$
    if $p$ is thread creation point and 
    $n_\mathit{sta}$ is the child thread's start node.

  \item $\textsc{ThJoins}$ is a parent--child relation over threads:
    $(p,n_\mathit{end}) \in \textsc{ThJoins}$ means $p$ is a
    thread join on a child thread with node $n_\mathit{end}$ as exit.

  \item 
    $(l, v) \in \textsc{IsLoad}$ means $l$ is a load of variable $v$.

  \item 
    $(s,v)\in \textsc{IsStore}$ means $s$ is a store to variable
    $v$.

  \item $\textsc{ReadsFrom}$ is obtained from the combination $i_c$ under test: $(l,s) \in \textsc{ReadsFrom}$ if the load $l$ is reading
    from the store $s$.

\end{asparaitem}
All these relations can be computed from the given set $\mathit{Gs}$
of control-flow graphs efficiently~\cite{Lengauer79}.  Furthermore,
they are defined over finite domains (sets of nodes or variables), which means
constraints built upon these relations are efficiently decidable.

\subsection{Deduction Rules for Checking Feasibility}

\begin{figure}
\vspace{1ex}
{\scriptsize
  \begin{equation}
    \label{pt:dom}
    \begin{mathprooftree}
      \AxiomC{
              $(m, n) \in \textsc{Dominates}
              \land (m, n) \in \textsc{NotReachableFrom}
              $
             }
      \UnaryInfC{$(m, n) \in \textsc{MHB}$}
    \end{mathprooftree}
  \end{equation}
  \vspace{1ex}

  \begin{equation}
    \label{pt:create}
    \begin{mathprooftree}
      \AxiomC{$(m, n_\mathit{sta}) \in \textsc{ThCreates}$}
      \UnaryInfC{$(m, n_\mathit{sta}) \in \textsc{MHB}$}
    \end{mathprooftree}
\hspace{3ex}
    \begin{mathprooftree}
      \AxiomC{$(m, n_\mathit{end}) \in \textsc{ThJoins}$}
      \UnaryInfC{$(n_\mathit{end}, m) \in \textsc{MHB}$}
    \end{mathprooftree}
  \end{equation}
  \vspace{1ex}

  \begin{equation}
    \label{pt:overwrite}
    \begin{mathprooftree}
      \AxiomC{
        \Shortstack{
            {
              $(l, s_1) \in \textsc{ReadsFrom} 
              \land (s_1, s_2) \in \textsc{MHB}$
            }
            {
              $\phantom{e} \land (l, v) \in \textsc{IsLoad}
              \land (s_1, v) \in \textsc{IsStore}$
            }
            {
              $\phantom{e} \land (s_2, v) \in \textsc{IsStore}$
            }
        } 
      } 
      \UnaryInfC{$(l, s_2) \in \textsc{MHB}$}
    \end{mathprooftree}
  \end{equation}
  \vspace{1ex}

  \begin{equation}
    \label{pt:trans}
    \begin{mathprooftree}
      \AxiomC{
        $(a, b) \in \textsc{MHB}
         \land (b, c) \in \textsc{MHB}$
      }
      \UnaryInfC{$(a, c) \in \textsc{MHB}$}
    \end{mathprooftree}
  \end{equation}
  \vspace{1ex}

  \begin{equation}
    \label{pt:mhb_nrf}
    \begin{mathprooftree}
      \AxiomC{
        $(a, b) \in \textsc{MHB}$
      }
      \UnaryInfC{$(a, b) \in \textsc{MustNotReadFrom}$}
    \end{mathprooftree}
  \end{equation}

  \begin{equation}
    \label{pt:l_after_s}
    \begin{mathprooftree}
      \AxiomC{
        \Shortstack{
          {
            $(l_1, s_1) \in \textsc{ReadsFrom}
            \land (l_1, s_2) \in \textsc{MHB}
            \land (s_2, l_2) \in \textsc{MHB}$
          }
          {
            $\phantom{e} \land (l_1, v) \in \textsc{IsLoad}
            \land (l_2, v) \in \textsc{IsLoad}
            \land (s_2, v) \in \textsc{IsStore}$
          }
        }
      }
      \UnaryInfC{$(l_2, s_1) \in \textsc{MustNotReadFrom}$}
    \end{mathprooftree}
  \end{equation}
}
  \caption{Rules used by our interference feasibility analysis.}
  \label{fig:rules}
\end{figure}


Figure~\ref{fig:rules} shows the deduction rules underlying our feasibility
analysis.  
%
%
If a contradiction is reached after applying the rules to the input
constraints, the interference combination is guaranteed to be
infeasible.
For brevity, we only present the intuition behind these rules.
Detailed proofs can be found in our supplementary material.

Rules~\ref{pt:dom}, \ref{pt:create}, and \ref{pt:overwrite} create the
\emph{must-happen-before} relation, $\textsc{MHB}$, where $(m, n) \in
\textsc{MHB}$ means node $m$ must happen before node $n$ under the
current interference combination $i_c$.
Rule~\ref{pt:trans} is simply the transitive property for the
must-happen-before relation.

First, if $m$ dominates $n$ in a CFG, since $m$
occurs before $n$ on \emph{all} program paths, $m$ must happen before
$n$ (Rule~\ref{pt:dom}).
We check if $n$ can reach $m$ to ensure that even if $m$ dominates $n$, $m$ can
never subsequently occur after $n$ (e.g., if $n$ is in a loop).
Similarly, since a thread cannot execute before it is created, or
after it terminates, $\textsc{ThCreates}$ and $\textsc{ThJoins}$ also
map directly to $\textsc{MHB}$ (Rule~\ref{pt:create}).

\begin{figure}
  \centering
  \begin{tikzpicture}
    \matrix (m) [matrix of nodes, 
      column sep=15mm,
      row sep=3mm,
      ampersand replacement=\&,
      nodes={draw, 
        line width=1pt,
        anchor=center, 
        text centered,
        rounded corners,
      },
        txt/.style={minimum width=2cm,anchor=center},
      ] {
         |[txt]| {\scriptsize $s_1$: \texttt{v = 5}}       
         \&
         \\
         |[txt]| {\scriptsize $s_2$: \texttt{v = 6}}       
         \& |[txt]| {\scriptsize $l$: \texttt{l = v}}       
         \\
      };
      \draw[-latex] (m-1-1.south) -- (m-2-1.north);
      \draw[-latex, dashed] (m-1-1.east) -- ($(m-2-2.west)+(0,0.5em)$); 
      \draw[-latex, dotted, red] (m-2-2.west) -- (m-2-1.east);
  \end{tikzpicture}

  \caption{Example: application of Rule~\ref{pt:overwrite}.} 
  \label{fig:overwrite}
\end{figure}
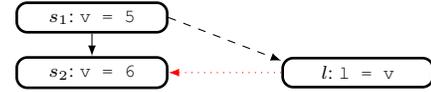

%
Rule~\ref{pt:overwrite} captures the scenario of two stores overwriting each
other as shown in Figure~\ref{fig:overwrite}. 
%
%
Here, one thread has stores $s_1$ and $s_2$, and a second  thread has one
load $l$.
$\Call{ReadsFrom}{l,s_1}$ is represented by the dashed edge (flow of
data) from $s_1$ to $l$.  
$\textsc{MHB}(s_1,s_2)$ is represented by the solid edge from $s_1$ to
$s_2$.
Given the two previous relations, the rule deduces the relation
$\textsc{MHB}(l,s_2)$, represented by the red dotted edge.
The implication is that for load $l$ to read from the first store $s_1$, $l$
must happen before the second store $s_2$.

The intuition behind this rule is that if $s_2$ executes before $l$, then $s_2$
would overwrite the value of $s_1$, making it impossible for $l$ to read
the value of $s_1$.  Note that this must-happen-before constraint is
\emph{only} considered for $i_c$, the current combination of interferences: it does
\emph{not} hold globally across all executions of the program.

Rule~\ref{pt:mhb_nrf} introduces the $\textsc{MustNotReadFrom}$ relation.
%
%
For a load store pair $(l,s) \in \textsc{MustNotReadFrom}$ if in the current
interference combination $l$ cannot read from $s$.
%

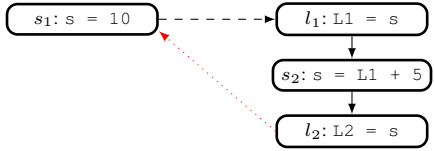
\begin{figure}
\vspace{1ex}
  \centering
  \begin{tikzpicture}
    \matrix (m) [matrix of nodes, 
      column sep=15mm,
      row sep=3mm,
      ampersand replacement=\&,
      nodes={draw, 
        line width=1pt,
        anchor=center, 
        text centered,
        rounded corners,
      },
        txt/.style={minimum width=2cm,anchor=center},
      ] {
         |[txt]| {\scriptsize $s_1$: \texttt{s = 10}}       
         \& |[txt]| {\scriptsize $l_1$: \texttt{L1 = s}}       
         \\
         \& |[txt]| {\scriptsize $s_2$: \texttt{s = L1 + 5}}     
         \\
         \& |[txt]| {\scriptsize $l_2$: \texttt{L2 = s}}     
         \\
      };
      \draw[-latex] (m-1-2.south) -- (m-2-2.north);
      \draw[-latex] (m-2-2.south) -- (m-3-2.north);
      \draw[-latex, dashed] (m-1-1.east) -- (m-1-2.west);
      \draw[-latex, dotted, red] (m-3-2.west) -- ($(m-1-1.east)+(0em,-0.5em)$);
  \end{tikzpicture}
  \caption{Example: application of Rule~\ref{pt:l_after_s}.}
  \label{fig:overw}
\end{figure}

Rule~\ref{pt:l_after_s} prevents a thread from reading an interference
after it has been over-written, shown in Figure~\ref{fig:overw}.
The first thread has a store $s_1$,
and the second thread has load $l_1$, store $s_2$, and then load $l_2$.
Again, $\textsc{MHB}$ relations are represented by solid edges,
$\textsc{ReadsFrom}(l_1,s_1)$ is represented by the dashed edge, and
$\textsc{MustNotReadFrom}(l_2,s_1)$ is represented by the red dotted
edge.

Conceptually, the rule captures the situation when a value is read
from an interference ($l_1$: \textcode{L1=s}), followed by a
modification of the same memory location that was loaded
($s_2$: \textcode{s=L1+5}), followed by a load of the same location
($l_2$: \textcode{L2=s}). Intuitively, since the interfering value was
just overwritten, it cannot be loaded again. Therefore, the pair
$(l_2,s_1)$ is added to $\textsc{MustNotReadFrom}$.

Finally, our constraint analysis does not try to identify all infeasible
combinations for efficiency reasons. However, the framework is generic enough
to allow new rules and other types of constraint solvers to be plugged in
easily to refine the approximation. 

\subsection{The Running Example}

\begin{figure}[t!]
  \centering
  \begin{tikzpicture}
    \matrix (m) [matrix of nodes, 
      column sep=15mm,
      row sep=3mm,
      ampersand replacement=\&,
      nodes={draw, 
        line width=1pt,
        anchor=center, 
        text centered,
        rounded corners,
      },
        txt/.style={minimum width=2cm,anchor=center},
      ] {
         |[txt]| {\scriptsize $l_4$: \texttt{x = 4}}       
         \& |[txt]| {\scriptsize $l_9$: \texttt{b1 = flag}}
         \\
         |[txt]| {\scriptsize $l_5$: \texttt{x = 5}}       
         \& |[txt]| {\scriptsize $l_{11}$: \texttt{t1 = x}}   
         \\
         |[txt]| {\scriptsize $l_{6}$: \texttt{flag = true}} 
         \& 
         \\
      };
      \draw[-latex] (m-1-1.south) -- (m-2-1.north);
      \draw[-latex] (m-2-1.south) -- (m-3-1.north);
      \draw[-latex] (m-1-2.south) -- (m-2-2.north);
      \draw[-latex, dashed] (m-3-1.east) -- ($(m-1-2.south)+ (-2.4em, 0)$);
      \draw[-latex, dashed] (m-1-1.east) -- ($(m-2-2.west)+(0,0.5em)$);
      \draw[-latex
            , dotted
            , red
           ]
      (m-2-2.west) -- (m-2-1.east);
      \path 
            (m-1-2.west)
        edge[-latex
             , dotted
             , red
            ]
($(m-3-1.north) + (3em, 0)$);
  \end{tikzpicture}
  \caption{Input and implied constraints for Figure~\ref{fig:ex01}.}
  \label{fig:ex01_constr}
\end{figure}
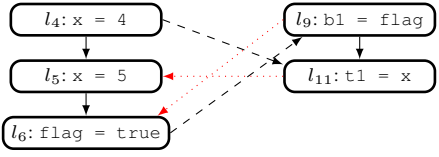

%
We revisit the example in Figure~\ref{fig:ex01} to illustrate our feasibility checking for one 
interference combination (Figure~\ref{fig:ex01_constr}). Our goal is to decide if
\textsc{ReadsFrom}$(l_9,l_6)$ and \textsc{ReadsFrom}$(l_{11},l_4)$ can
co-exist.
Initially, our constraint system would have the solid edges from the
$\textsc{MHB}$ relations, which represent the program-order
constraints, and the dashed edges from the $\textsc{ReadsFrom}$
relations, which represent the current interference combination $i_c$.

First, we can deduce $\textsc{MHB}(l_{11}, l_5)$ by applying
Rule~\ref{pt:overwrite}: if $l_{11}$ does not happen before $l_5$, 
$l_5$ would overwrite the value of \textcode{x}, preventing $l_{11}$
from reading from $l_4$.
This deduced $\textsc{MHB}$ relation is represented by the red dotted
edge in the figure.

Next, we can deduce a must-happen-before relation between $l_9$ and
$l_6$ by applying Rule~\ref{pt:trans} twice.  That is,
$\Call{MHB}{l_9,l_{11}} \wedge \Call{MHB}{l_{11},l_5}$ implies
$\Call{MHB}{l_9,l_5}$, followed by  $\Call{MHB}{l_9,l_5} \wedge
\Call{MHB}{l_5,\allowbreak l_6}$ implies $\Call{MHB}{l_9,l_6}$.  The result is
represented by the red dotted edge from $l_9$ to $l_6$.

At this point, we have a contradiction: since \textcode{b1=flag}
must-happen-before \textcode{flag=true}, \textcode{b1} cannot read
the value of \textcode{true} (Rule~\ref{pt:mhb_nrf}). So, this
interference combination is proved to be infeasible. (There are more 
implied edges in Figure~\ref{fig:ex01_constr}; for clarity, we show
only those relevant to the check.)


\section{Optimizations with Clustering and Pruning}
\label{sec:pruning}

To reduce the number of interference combinations, we apply
dependency-based clustering analysis and property-directed pruning.
Consider the program in Figure~\ref{fig:ex03}: the \textcode{main}
thread creates two children in the function \textcode{thr} with
arguments \textcode{5} and \textcode{10}, respectively.  The
\textcode{thr} function performs a store to \textcode{x} (Line~5)
based on the value passed as an argument (\textcode{v}).  At the load
of \textcode{x} in the \textcode{thr} function, the value may come
from the initial value \textcode{0}, from the \textcode{main} thread
(Line~12), or from the other thread \textcode{thr} (Line~5).  This
results in three combinations of loads in \textcode{thr} to be tested
on every iteration.

\begin{figure}
\vspace{1ex}
  \centering
  \begin{minipage}{0.35\linewidth}
    \begin{lstlisting}[language=C, numbers=left, frame=single, name=motiv3]
int x = 0;
void thr(int v) {
  int t1 = 5 * v;
  int t2 = x;
  x = t1 + t2;
  if (t1 < 0) 
    ERROR!;
}
    \end{lstlisting}
  \end{minipage} \hspace{0.1\linewidth}
  \begin{minipage}{0.425\linewidth}
	  \begin{lstlisting}[language=C, numbers=left, frame=single, name=motiv3]
int main() {
  thread_create(thr,5);
  thread_create(thr,10);
  x = 1;
  thread_exit(0);
}

    \end{lstlisting}
  \end{minipage}
  \caption{Example: property directed redundancy pruning.}
  \label{fig:ex03}
\end{figure}

However, the reachability of
\textcode{ERROR!} does \emph{not} depend on the value loaded from \textcode{x}, since 
 the error condition ($\textcode{t1}<0$) only
depends on the argument passed to \textcode{thr}.  As such, the load of
\textcode{x} is immaterial to the property.  We can formally capture
this notion of immateriality using \emph{control and data
dependencies}~\cite{Ferrante87}.

Intuitively, a statement $s$ is data dependent on $t$ if the value of
$t$ may affect the computation of $s$. For example, in
Figure~\ref{fig:ex03}, the statement \textcode{t1=5*v} is data
dependent on the input parameter \textcode{v}.
On the other hand, a statement $l$ is control dependent on $m$ if the
execution of $m$ affects the reachability of $l$. For example, the
\textcode{ERROR!}  statement in Figure~\ref{fig:ex03} is
control-dependent on the evaluation of the predicate \textcode{t1<0}.

%
%
The composition of the control- and data-dependency relations is the
\emph{program dependence graph}~\cite{Ferrante87}.
Note that in concurrent programs, the dependency graph may
span across multiple threads, due to the flow of data from shared memory writes
to reads.

Next, we show two applications of the program dependence graph for
optimizing our overall algorithm.  
%

\subsection{Property-guided Pruning}

\begin{figure}
\vspace{1ex}
    \centering
    \def\svgwidth{\columnwidth}
    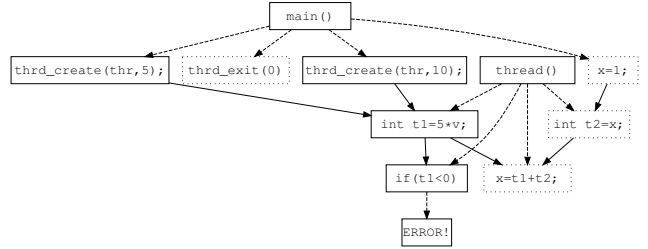
    \vspace{-1.5em}
  \caption{The program dependence graph for Figure~\ref{fig:ex03}.}
  \label{fig:pdg}
\end{figure}

%
First, we create the \emph{backward slice} on every property in the
program.
The backward slice with respect to a property $s$ contains all the
statements involved in the computation of $s$ (Theorem 2.2
\cite{Kenn01}).
%
%
As an example, the program dependence graph for Figure~\ref{fig:ex03}
is shown in Figure~\ref{fig:pdg}. Dashed edges are control
dependencies and solid edges are data dependencies. The backward
slice on the \textcode{ERROR!} statement is also shown: the dotted
nodes are nodes not contained in the slice. 
All computations involving \textcode{x} can be ignored, since the
slice shows that they are irrelevant to the property being verified.

During our analysis, the transfer function of a statement not on the
backward slice is the \emph{identity}. 
%
%
%
And any load not on the backward slice is ignored when computing
interference combinations.

\subsection{Dependency-guided Clustering}

Second, during the generation of combinations of interferences, we do not
always consider the Cartesian product across all sets of loads.
Instead, we group loads together to form \emph{cluster} and only generate
interference combinations within each cluster.

\begin{figure}
\vspace{1ex}
\centering
  \begin{minipage}{0.35\linewidth}
    \begin{lstlisting}[language=C, numbers=left, frame=single, name=prop]
int x = 0;
int y = 0;
void thread1() {
  x = 1;
  y = 1;
}
    \end{lstlisting}
  \end{minipage}\hspace{0.15\linewidth}
  \begin{minipage}{0.35\linewidth}
    \begin{lstlisting}[language=C, numbers=left, frame=single, name=prop]
void thread2() {
  int t1 = x;
  int t2 = y;
  assert(x >= 0);
  assert(y >= 0);
}
    \end{lstlisting}
  \end{minipage}
  \caption{Example: dependency guided clustering.}
  \label{fig:prop_ex}
\end{figure}

Consider the program in Figure~\ref{fig:prop_ex}. Initially,
\textcode{x} and \textcode{y} are zero; the first thread sets them to one, and
the second thread checks the property that they are both greater than or equal
to zero. The backward slice on \textcode{assert(x>=0)} contains 
lines 4, 8, and 10.
The backward slice on \textcode{assert(y>=0)} contains 
lines 5, 9, and 11.
%
%
Without optimization, in Algorithm~\ref{alg:constr_ai}, the loads on
\textcode{x} and \textcode{y} both have two potential environments to
read from: the interfering store and the environment within the
thread. In total, there are $2*2=4$ 
combinations leading to four abstract interpreter executions.



The backward slices on properties in the program form disjoint subgraphs; e.g.,
a graph with the operations on \textcode{x} and those on \textcode{y}.
%
The interference combinations in the subgraphs can be considered independently
requiring only $\mathit{max}(2,2) = 2$ interpreter executions.




\section{Experiments}
\label{sec:exp}

We implemented our method in a software tool named \textsc{Watts},
designed for verifying multithreaded programs represented in the LLVM intermediate
language. All experiments were performed on C programs written using
POSIX threads.  We used the Apron library~\cite{Jeannet09} for
implementing the sequential analyzer over interval and octagon
abstract domains, and the Datalog solver in Z3
($\mu${Z}~\cite{Hoder11}) for solving the causality constraints.

We evaluated \textsc{Watts} on two sets of benchmark programs. The
first set consists of some multithreaded programs from
SVCOMP~\cite{svcomp15}.  The second set consists of Linux
device drivers from~\cite{LinuxISR} and~\cite{Farzan12}.  In all
benchmark programs, the reachability properties are expressed in the
form of embedded assertions.  Table~\ref{tbl:prog_sum} shows the
characteristics of these programs, including the name, the number of
lines of code (LoC), the number of threads, and the number of
assertions.  In total, our benchmarks have 26,309 lines of code and
10,078 assertions.  For the device driver benchmarks, in particular,
since assertions are not included in the original source code, we
manually added these assertions.  We performed all experiments on a
computer with 8 GB RAM and a 2.60 GHz CPU.

\begin{table}[t!]
\caption{Statistics of the benchmarks in our experiments.}
\label{tbl:prog_sum}
	\centering
  \resizebox*{\linewidth}{!}{%
\addtolength{\tabcolsep}{+3pt}
	\begin{tabular}{lcccc}
		\toprule
		Name            & LoC  & Threads & Properties & Source\\
		\midrule
		thread01        & 29   & 3       & 1      & created         \\
		create01        & 24   & 2       & 1      & created     \\
		create02        & 28   & 2       & 1      & created     \\
		sync01          & 38   & 3       & 1      & \cite{svcomp15}     \\
		sync02          & 36   & 3       & 1      & \cite{svcomp15}     \\
		intra01         & 41   & 3       & 1      & created     \\
		dekker1         & 65   & 3       & 1      & \cite{svcomp15}     \\
		fk2012          & 88   & 3       & 1      & \cite{Farzan12}, added asserts     \\
		\midrule
		keybISR         & 62   & 3       & 2      & \cite{LinuxISR}     \\
		ib700\_01       & 346  & 3       & 1      & \cite{Farzan12}, added asserts     \\
		ib700\_02       & 466  & 23      & 1      & \cite{Farzan12}, added asserts    \\
		ib700\_03       & 587  & 41      & 81     & \cite{Farzan12}, added asserts    \\
		i8xxtco\_01     & 735  & 3       & 1      & \cite{Farzan12}, added asserts    \\
		i8xxtco\_02     & 901  & 22      & 1      & \cite{Farzan12}, added asserts    \\
		i8xxtco\_03     & 1027 & 42      & 103    & \cite{Farzan12}, added asserts    \\
		machz\_01       & 667  & 8       & 1      & \cite{Farzan12}, added asserts    \\
		machz\_02       & 795  & 29      & 1      & \cite{Farzan12}, added asserts    \\
		machz\_03       & 881  & 41      & 83     & \cite{Farzan12}, added asserts    \\
		mix\_01         & 457  & 12      & 1      & \cite{Farzan12}, added asserts    \\
		mix\_02         & 580  & 31      & 62     & \cite{Farzan12}, added asserts    \\
		pcwd\_01        & 1197 & 8       & 1      & \cite{Farzan12}, added asserts    \\
		pcwd\_02        & 1405 & 41      & 81     & \cite{Farzan12}, added asserts    \\
		sbc\_01         & 686  & 24      & 1      & \cite{Farzan12}, added asserts    \\
		sc1200\_01      & 715  & 24      & 1      & \cite{Farzan12}, added asserts    \\
		sc1200\_02      & 768  & 31      & 93     & \cite{Farzan12}, added asserts    \\
		smsc\_01        & 904  & 12      & 1      & \cite{Farzan12}, added asserts    \\
		smsc\_02        & 931  & 12      & 24     & \cite{Farzan12}, added asserts     \\
		sc520\_01       & 806  & 4       & 1      & \cite{Farzan12}, added asserts     \\
		sc520\_02       & 880  & 41      & 81     & \cite{Farzan12}, added asserts     \\
		wfwdt\_01       & 777  & 4       & 1      & \cite{Farzan12}, added asserts     \\
		wfwdt\_02       & 907  & 51      & 101    & \cite{Farzan12}, added asserts     \\
		wdt             & 1023 & 31      & 1      & \cite{Farzan12}, added asserts     \\
		wdt977\_01      & 867  & 16      & 1      & \cite{Farzan12}, added asserts     \\
		wdt977\_02      & 877  & 31      & 92     & \cite{Farzan12}, added asserts     \\
		wdt\_pci        & 1133 & 31      & 1      & \cite{Farzan12}, added asserts     \\
		wdt\_pci02      & 1165 & 31      & 122    & \cite{Farzan12}, added asserts     \\
		pcwdpci\_01~~~~~~~& 1363 & 64      & 128    & \cite{Farzan12}, added asserts   \\
		\bottomrule
	\end{tabular}
\addtolength{\tabcolsep}{-3pt}
}
\end{table}

Although we used the benchmarks from~\cite{Farzan12},
the verification problem targeted by our method is significantly
different.  \textsc{Duet} assumes each device driver is a parametric
program, whereas our method analyzes programs with a finite number of
threads.  As shown in Section~\ref{sec:motiv}, our method, using a set
of control-flow graphs as opposed to a monolithic data-flow graph, is
often more accurate.
During experiments, we ran both \textsc{Watts} and \textsc{Duet} on
all benchmarks with our assertions.  \textsc{Watts} verified 548 more
properties than \textsc{Duet}, whereas \textsc{Duet} did not verify
any property not verified by \textsc{Watts}.  The result shows that
\textsc{Duet}'s abstraction for infinite threads leads to loss of
precision.
Therefore, in the remainder of this section, we do not directly
compare \textsc{Watts} with \textsc{Duet}.

Instead, we focus on comparing our method with the prior 
thread-modular approaches~\cite{Ferrara08, Mine11, Mine12, Mine14}.
For evaluation purposes, we implemented both methods in
\textsc{Watts}: the flow-insensitive analysis of
Algorithm~\ref{alg:constr_ai} and the flow-insensitive analysis of
Algorithm~\ref{alg:tm_ai}.

Table~\ref{tbl:box_result} shows the results of comparing
Algorithm~\ref{alg:constr_ai} and Algorithm~\ref{alg:tm_ai} in the
interval abstract domain.  Column~1 shows the name of each benchmark.
Columns 2--3 show the result of running Algorithm~\ref{alg:tm_ai}.
Columns 4--5 show the result of running Algorithm~\ref{alg:constr_ai}
without using the feasibility checking.  Columns 6--7 show the result
with the feasibility checking.  Columns 8--9 show the result with
clustering/pruning optimizations.  For each test case, Tm.\ is the run
time in seconds and Verif.\ is the number of verified properties.  The
last row shows the sum of all columns.

\begin{table}[!t]
\caption{Experimental results in the interval domain.}
\label{tbl:box_result}
  \centering
  \resizebox*{\linewidth}{!}{%
  \addtolength{\tabcolsep}{-3pt}
  \begin{tabular}{l*{9}{c}}
    \toprule
    & \multicolumn{2}{c}{Flow-insensitive}
    & \multicolumn{2}{c}{Flow-sensitive}
    & \multicolumn{2}{c}{F.-s. + Const.}
    & \multicolumn{2}{c}{F.-s. + Opt.} \\
    \cmidrule(r){2-3}
    \cmidrule(r){4-5}
    \cmidrule(r){6-7}
    \cmidrule(r){8-9}
    Name     
		& Tm.\ (s) & Verif.  
		& Tm.\ (s) & Verif.  
		& Tm.\ (s) & Verif.  
		& Tm.\ (s) & Verif.  \\
    \midrule
 thread01    & 0.03   & 0  & 0.05   & 0   & 0.05   & 1    & 0.09    & 1 \\
 create01    & 0.02   & 0  & 0.03   & 0   & 0.04   & 1    & 0.07    & 1 \\
 create02    & 0.03   & 0  & 0.03   & 0   & 0.03   & 1    & 0.07    & 1 \\
 sync01      & 0.04   & 0  & 0.05   & 1   & 0.06   & 1    & 0.07    & 1 \\
 sync02      & 0.04   & 0  & 0.06   & 0   & 0.07   & 1    & 0.07    & 1 \\
 intra01     & 0.03   & 0  & 0.03   & 0   & 0.03   & 1    & 0.08    & 1 \\
 dekker1     & 0.14   & 0  & 9.81   & 0   & 2.10   & 1    & 0.75    & 1 \\
 fk2012      & 0.10   & 0  & 0.25   & 0   & 0.25   & 1    & 0.18    & 1 \\
 \midrule    
 keybISR     & 0.05   & 0  & 0.15   & 0   & 0.14   & 2    & 0.12    & 2 \\
 ib700\_01   & 0.09   & 0  & 0.10   & 0   & 0.10   & 1    & 0.13    & 1 \\
 ib700\_02   & 1.17   & 0  & 0.88   & 0   & 0.95   & 1    & 1.03   & 1 \\
 ib700\_03   & 33.46  & 0  & 40.95  & 40  & 36.95  & 81   & 37.81  & 81 \\
 i8xxtco\_01 & 0.15   & 0  & 0.13   & 0   & 0.13   & 1    & 0.22    & 1 \\
 i8xxtco\_02 & 1.34   & 0  & 0.96   & 0   & 1.02   & 1    & 1.24   & 1 \\
 i8xxtco\_03 & 38.07  & 18 & 50.78  & 61  & 47.90  & 103  & 55.24  & 103 \\
 machz\_01   & 0.21   & 0  & 0.18   & 0   & 0.18   & 1    & 0.29    & 1 \\
 machz\_02   & 0.97   & 0  & 0.69   & 0   & 0.76   & 1    & 0.94    & 1 \\
 machz\_03   & 41.30  & 0  & 74.32  & 42  & 153.50 & 83   & 118.25 & 83 \\
 mix\_01     & 0.24   & 0  & 0.19   & 0   & 0.20   & 1    & 0.29    & 1 \\
 mix\_02     & 12.42  & 1  & 15.22  & 31  & 13.24  & 62   & 15.28  & 62 \\
 pcwd\_01    & 0.25   & 0  & 0.21   & 0   & 0.21   & 1    & 0.32    & 1 \\
 pcwd\_02    & 33.12  & 0  & 41.57  & 40  & 33.77  & 81   & 38.23  & 81 \\
 sbc\_01     & 0.60   & 0  & 0.73   & 0   & 1.09   & 1    & 0.57    & 1 \\
 sc1200\_01  & 0.53   & 0  & 0.62   & 0   & 0.47   & 1    & 0.54    & 1 \\
 sc1200\_02  & 70.46  & 0  & 119.24 & 62  & 161.00 & 93   & 122.48 & 93 \\
 smsc\_01    & 0.35   & 0  & 0.32   & 0   & 0.40   & 1    & 0.51    & 1 \\
 smsc\_02    & 3.73   & 0  & 7.27   & 1   & 15.39  & 24   & 6.12   & 24 \\
 sc520\_01   & 0.64   & 0  & 1.23   & 0   & 0.73   & 1    & 0.72    & 1 \\
 sc520\_02   & 50.87  & 0  & 81.27  & 39  & 65.95  & 81   & 46.95  & 81 \\
 wfwdt\_01   & 0.61   & 0  & 1.20   & 0   & 0.71   & 1    & 0.70    & 1 \\
 wfwdt\_02   & 94.54  & 0  & 148.32 & 0   & 118.39 & 101  & 83.91  & 101 \\
 wdt         & 0.71   & 0  & 0.49   & 0   & 0.55   & 1    & 0.69    & 1 \\
 wdt977\_01  & 0.57   & 0  & 0.44   & 0   & 0.49   & 1    & 0.65    & 1 \\
 wdt977\_02  & 51.86  & 0  & 58.92  & 32  & 86.16  & 93   & 92.01  & 93 \\
 wdt\_pci    & 0.79   & 0  & 0.55   & 0   & 0.61   & 1    & 0.77    & 1 \\
 wdt\_pci02  & 75.14  & 1  & 114.55 & 31  & 100.12 & 122  & 110.33 & 122 \\
 pcwdpci\_01 & 91.10  & 18 & 115.82 & 72  & 136.13 & 128  & 109.02 & 128 \\
\midrule
\textbf{Total}   & 605.77 & 38 & 887.61 & 452 & 979.87 & 1,078 & 846.74 & 1,078 \\
    \bottomrule
  \end{tabular}
  \addtolength{\tabcolsep}{+3pt}
}
\end{table}

Compared to the flow-insensitive approach (Columns 2--3), our
baseline flow-sensitive method (Columns 4--5) can already achieve a
12x increase in the number of verified properties (from 38 to 452)
without employing the lightweight constraint-based feasibility
checking.  This demonstrates the benefits of delaying the join
operation across threads.  Furthermore, the significant increase in
accuracy comes at the modest 1.5x increase in run time.

With the constraint-based feasibility checking, a more significant
improvement can be observed (Columns 6--7): there is a 28x increase in
the number of verified properties (from 38 to 1,078) compared to the
prior flow-insensitive approach.  Furthermore, the large increase in
accuracy comes with only an 1.6x increase in run time. 

Finally, with the optimizations from
Section~\ref{sec:pruning}, our method improves further
(Columns 8--9). Compared to the prior flow-insensitive approach
(Columns 2--3), our method only has a 1.4x increase in the runtime
overhead but with a 28x increase in number of verified
properties.  Compared to the version of our method without
optimizations (Columns 6--7), the version with optimization finishes
the entire analysis 1.4x faster. 
Additionally, the optimized version finishes slightly \emph{faster} than the
non-constraint based approach (Columns 4--5) while able to verify 2.4x as many
properties.

%

Note that across all experiments, the number of verified properties are
strictly increasing: e.g., the flow-sensitive approach with optimizations
verifies all the properties of the flow-insensitive approach and more. At most
we were able to verify 1,078 properties. Those we missed largely
were due to cross-thread synchronization which was not captured by our
constraint analysis.

In addition to the results in Table~\ref{tbl:box_result}, we also
performed experiments using the octagon abstract domain.  We observed
little increase in accuracy as a result of this change, indicating
that the properties being verified are mostly on inter-thread
concurrency control behavior, and therefore a more sophisticated
representation of numerical relations over the program variables does
not offer more advantages.  For brevity, we omit the result table for
the octagon domain.

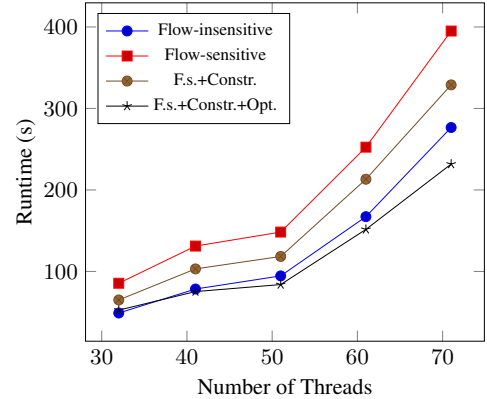
\begin{figure}
\vspace{1ex}
  \centering
  \resizebox*{0.75\linewidth}{!}{%
  \begin{tikzpicture}
    \begin{axis}[
      scale=0.35,
      width=\textwidth,
      yticklabel style={align=right,inner sep=0pt,xshift=-0.1cm},
      xlabel={Number of Threads},
      ylabel={Runtime (s)},
      legend pos=north west,
      legend entries={{\scriptsize Flow-insensitive}
                      , {\scriptsize Flow-sensitive}
                      , {\scriptsize F.s.+Constr.}
                      , {\scriptsize F.s.+Constr.+Opt.}
                      }
    ]
      \addplot table {nconstr.dat};
      \addplot table {comb.dat};
      \addplot table {constr.dat};
      \addplot table {slice.dat};
    \end{axis}
  \end{tikzpicture}%
  }
  \caption{Runtime overhead versus number of threads.}
  \label{fig:growth}
\end{figure}

In the past, introducing flow-sensitivity to static analysis often
results in scalability issues (e.g., \cite{Jeannet13}); however, this
is not the case for our method.  Figure~\ref{fig:growth} shows our
experiments on a parametrized program named \textit{i8xx\_tco}, where
the run time of our method grows only moderately with the increase in
program size.  Here, the $x$-axis is the number of threads of the
program and the $y$-axis is the run time.
The optimized  method has slightly \emph{lower} runtime than the
least accurate flow-insensitive approach.
%
%
Furthermore, our method enjoys an almost linear growth in the
execution time, indicating it is more scalable than the other methods.

\section{Related Work}
\label{sec:related}

There is a large body of work on the static analysis and formal
verification of multithreaded programs, but none of these existing
methods can obtain flow sensitivity in thread-modular analysis with a
reasonable run-time cost.  For brevity, we discuss only those that are
most relevant to our new method.  The interested reader can see
Rinard~\cite{Rinard01} for a survey of early work.



Thread-modular abstract interpretation was introduced by
Ferrara~\cite{Ferrara08} and Min{\'e}~\cite{Mine11, Mine12}. As 
shown, their approaches eagerly joined interferences and
considered them flow-insensitively, thus introducing
inaccuracies.  Our method avoids such drawbacks.
Ferrara~\cite{Ferrara08} also introduced models designed specific for
the Java memory model to remove certain types of infeasible
interferences in an \emph{ad hoc} fashion.  Our constraint-based
feasibility checking, in contrast, is more general and systematic, and
can handle transitive must-happen-before constraints as well as other
constraints both within and across threads.

Min{\'e}~\cite{Mine14} introduced an extension to their prior
thread-modular analysis to compute \emph{relational}
interferences. This allows for relations between variables to be
maintained across threads, thereby bringing more accuracy than using
\emph{non-relational} interferences.  However, as we have explained
earlier, this technique is orthogonal and complementary to our new
method.

Farzan and Kincaid~\cite{Farzan12} introduced a method to iteratively
construct a monolithic data-flow graph for a concurrent
program. However, their technique, as well as similar methods designed
for parametric programs~\cite{KaiserKW10,TorreMP10}, targets the
problem of verifying properties in a concurrent program with an
unbounded number of threads.  As we have shown earlier, our new method
is often significantly more accurate than these existing methods.

Thread-modular approaches have been applied to model
checking~\cite{Flan03,HenzingerJMQ03} and symbolic
analysis~\cite{SinhaW10,SinhaW11}.
There are also works on verifying concurrent software using
abstraction and stateless model
checking~\cite{Wang08b,Yang09,Wang11,KusanoW14}.
However, these approaches in general are either heavyweight or
under-approximative, and therefore are complementary to our
abstract-interpretation based approach.



\section{Conclusions}
\label{sec:conc}

We have presented a flow-sensitive method for composing standard
abstract interpreters to form a more accurate thread-modular analysis
procedure for concurrent programs.  Our method relies on constructing
and solving a system of happens-before constraints to decide the
feasibility of inter-thread interference combinations.  We also use
clustering and pruning to reduce the run-time overhead of our
analysis.  We have implemented our method in a software tool and
evaluated it on a large set of multithreaded C programs.  Our
experimental results show that the new method can significantly
increase the accuracy of the thread-modular analysis while maintaining
a modest run-time overhead.

\section{Acknowledgments}

This work was primarily supported by
the NSF under grants CCF-1149454, CCF-1405697, and CCF-1500024.
Partial support was provided by
the ONR under grant N00014-13-1-0527.  
%

\newpage\clearpage
\bibliographystyle{abbrv}
\bibliography{mk}

\end{document}